\documentclass{article}

\usepackage{graphicx} 
\usepackage[a4paper, total={6in,8in}]{geometry}

\usepackage[numbers]{natbib}
\usepackage{authblk}
\usepackage{amsmath}
\usepackage{amsfonts}
\usepackage{xcolor}
\usepackage{wrapfig}

\usepackage[inkscapelatex=false]{svg}

\newcommand{\R}{\mathbb{R}}
\DeclareMathOperator{\dtm}{dtm}

\title{Topological Data Analysis for Unsupervised Feature Selection in Large Scale Spatial Omics Data Sets}

\author[1,2*]{James Boyle}
\author[3]{Gregory Hamm}
\author[4,5]{Eleanor Williams}
\author[2]{Robin JG Hartman}
\author[2]{Magnus S\"{o}derburg}
\author[2]{Ian Henry}
\author[2]{Michael Casey}

\affil[1]{Mathematical Institute, University of Oxford, Oxford, UK}
\affil[2]{Data Science and AI, Translational Science \& Experimental Medicine, Research and Early
Development, Cardiovascular, Renal and Metabolism, Biopharmaceuticals R\&D, AstraZeneca, Cambridge, UK}
\affil[3]{Integrated Bioanalysis, Clinical Pharmacology and Safety Sciences, Biopharmaceuticals
R\&D, AstraZeneca, Cambridge, UK}
\affil[4]{Cambridge Stem Cell Institute, University of Cambridge, Cambridge, UK}
\affil[5]{Predictive AI and Data, Clinical Pathology and Safety Sciences, Biopharmaceuticals R\&D, AstraZeneca, Cambridge, UK}
\affil[6]{Cardiovascular Renal and Metabolism Pathology, Clinical Pharmacology and Safety
Sciences, Biopharmaceuticals R\&D, AstraZeneca, Cambridge, UK}

\affil[*]{Correspondence to james.boyle@maths.ox.ac.uk}

\date{}

\begin{document}

\maketitle

\abstract{
Spatial transcriptomics studies are becoming increasingly large and commonplace, necessitating  simultaneous analysis of a large number of spatially resolved variables. Correspondingly, a diverse range of methodologies have been proposed to compare the spatial expression structure of genes. Here, we apply persistent homology, a method from topological data analysis, to produce a continuous quantification of spatial structure in a given gene's expression, and show how this can be used for downstream tasks such as spatially variable gene identification. We explore the unique advantages of topology for this task, deriving biologically meaningful insights into kidney disease and myocardial infarction using public spatial transcriptomics data. We also show how the non-parametric nature of homology enables our methodology to extend naturally to other spatial omics modalities, demonstrating this on a spatial metabolomics sample. Our work showcases the advantages of using a continuous quantification of spatial structure over \textit{p}-value based approaches to SVG identification, the potential for developing unified methods for the analysis of different spatial omics modalities, and the utility of persistent homology in big data applications.
}

\section{Introduction}
Spatial transcriptomics experiments measure gene expression in 2-dimensional space, up to the resolution of a supra-cellular well, cell, or subcellular location \cite{moses2022museum, marx2021method}. A common task when analysing spatial transcriptomics data is to identify genes that exhibit spatial structure in their expression, commonly referred to as \emph{Spatially Variable Genes} (SVGs) \cite{svg_methods_benchmarking, spatialde_paper}. 

SVG identification methods are typically based on null hypothesis rejection, with the null assuming no dependence of expression on spatial location \cite{spatialde_paper, sparkx_paper}. Here, as in \cite{sepal_paper}, we consider the degree of spatial variability, not a binary presence/absence, quantifying ``spatial structure" as a continuous descriptive quantity. This quantification of spatial structure can be used for identifying SVGs, but, as we show, it can also be useful for other analyses that involve comparing the level of spatial expression structure between different genes or tissue samples.

When quantifying spatial structure, we want our metric to be robust to the reasonable amounts of noise, tissue deformation and variations in tissue morphology that are inherent to spatial transcriptomics experiments. In addition, the metric should also be able to detect a very broad range of spatial structures. In other words, we wish to quantify the level of spatial structure in gene expression in a way that is minimally sensitive to the specific geometry or coordinate system of the tissue section, whilst being comprehensive as to what counts as spatial structure. These considerations lead very naturally to the mathematical field of topology and the tools of topological data analysis, which aim to provide coordinate-free characterisations of spatial organisation that are preserved under continuous deformations.

We present an approach based on persistent homology (PH) \cite{ph_roadmap, wasserman} that quantifies spatial variability via the topological activity in a gene’s expression pattern \cite{carlsson2009topology}. Loosely speaking, our methodology is based on quantifying the number and significance of `hotspots’ of expression, taking the presence of one or more distinct regions of differential gene expression as indicative of spatial structure. Via this highly generalised notion of spatial structure, we are able to detect a broad range of spatial patterns in a way that is robust to variance in tissue morphology. Moreover, using topology we are able to avoid restricting assumptions about the statistical distribution of gene expression or gene count data. This enables our methodology to be naturally extended to other spatial omics modalities, which we demonstrate by analysing a mass spectrometry imaging sample \cite{msi_overview}. 

We explore the capabilities of our persistent homology for SVG identification by using our approach to analyse spatial transcriptomics data from kidney disease and myocardial infarction samples. We also show how our topological quantification of spatial structure can be used to automatically identify genes which show a difference in spatial expression between Acute Kidney Injury and Chronic Kidney Disease samples. We illustrate the generalisability of persistent homology by using our methodology to identify spatially variable metabolites in a spatial metabolomics sample. 

In order to assess whether persistent homology brings new capabilities to the task of SVG identification, we also compare results from our method with those obtained using a range of other popular SVG identification techniques. We find that persistent homology detects a broader range of spatial structures, and produces more consistent results across different samples and biological settings. In comparison to Sepal, the only other SVG identification method the authors are aware of that performs SVG identification via a continuous quantification of spatial variability, we find that our topology based score identifies novel patterns of spatial structure, and is more effective for additional analytical tasks. 

Our methodology also has some novelty as an application of persistent homology to the automatic analysis of a large number of spatially resolved variables. Persistent homology has previously been used for analysing spatial structure in spatial transcriptomics data and other biological settings \cite{rizvi2017single, rabadan2019topological, benjamin2022multiscale}, but as far as the present authors are aware, the present work is more unusual as a ``big data" application of persistent homology, in which persistent homology is used to automatically perform some analytical task on a large number of spatially resolved variables, the output of which does not involve any manual inspection of the persistent homology outputs. In other words, our pipeline could be comfortably used by a practitioner without any knowledge of persistent homology or topological data analysis. We thus hope this work highlights not just the utility of persistent homology for analysing full transcriptome spatial transcriptomics data sets, but also the potential of persistent homology for use in other high dimensional settings.

We find our spatial structure score most effective as an unsupervised exploratory tool. That is, given spatial transcriptomics data on a large number of genes, we find that persistent homology effectively triages the data down to a smaller number of genes which, for example, display notable spatial structure, which can then be subjected to further analysis.

The rest of the paper is organised as follows. We first provide a non-technical outline of the process of going from the output of a spatial transcriptomics experiment to a list of SVGs. We then fill in the technical details, highlighting the special considerations needed when applying persistent homology to SVG identification. We then evaluate the capabilities of our spatial structure score for SVG identification by applying our methodology to spatial transcriptomics data from kidney disease and myocardial infarction samples, before showing how our persistent homology score can be used for other analytical tasks, and can be extended to other spatial omics modalities.

\section{Method Overview}\label{method_overview}
We take as input the spatially resolved expression of a large number of genes over a fixed set of co-ordinates. For each gene, we use persistent homology to compute a single number, the \emph{Coefficient of Spatial Structure} (CoSS), that quantifies the amount of spatial structure in that gene's expression pattern (figure \ref{persist_outline}a). In this section we provide a non-technical overview of how we compute the CoSS for a single gene (figure \ref{persist_outline}). For for a full description see \emph{Methods}.

Roughly speaking, the CoSS for a gene is computed by looking at the number and significance of regions of substantially higher or lower gene expression than the surrounding tissue. 

First we compute a smoothed version of the gene’s expression (figure \ref{persist_outline}b, \cite{wasserman}). Intuitively, for SVGs we would expect the surface plot of this smoothed expression to be more `hilly'. We then look at \emph{level sets} of the smoothed expression - regions of the underlying tissue where the smoothed expression exceeds a specified threshold (figure \ref{persist_outline}e). We consider these level sets over a continuous range of thresholds, from the maximum value of the smoothed expression down to zero. Regions of higher expression will be present in the level set at higher thresholds than regions of lower expression. As the threshold varies from its maximum down to zero, a region of the tissue that has significantly higher expression than the surrounding tissue will appear (in the language of persistent homlogy, it will be `born') as a disconnected component in the level set at a high threshold, and will only merge with the rest of the tissue (`die') at a low threshold, when the surrounding tissue appears in the level set \cite{wasserman}. By contrast, a region with expression only slightly higher than the surrounding tissue will die shortly after it is born. By looking at the `lifetime' of a hotspot - the difference between its birth and death threshold - we can measure the significance of the hotspot, and distinguish spatial signal from spatial noise.

The information from this level set analysis is neatly summarised in a \emph{barcode diagram} (figure \ref{persist_outline}c, \cite{wasserman}), consisting of a bar for each hotspot spanning from its birth to death threshold. This is a coarse but tractable summary of the spatial structure of the gene's expression. The CoSS is computed as the $L^2$-norm of the barcode, i.e.~by summing up the squared lengths of each bar and taking the square root \cite{carriere2015stable, adams2017persistence}. The CoSS is thus a summary of the number and lifetimes of a gene’s expression hotspots. This metric can then be used for downstream tasks that involve comparing the spatial expression structure of different genes. 

\subsubsection*{Spatially Variable Gene Identification}
\begin{figure}[t]
    \centering
    \includegraphics[width=\textwidth]{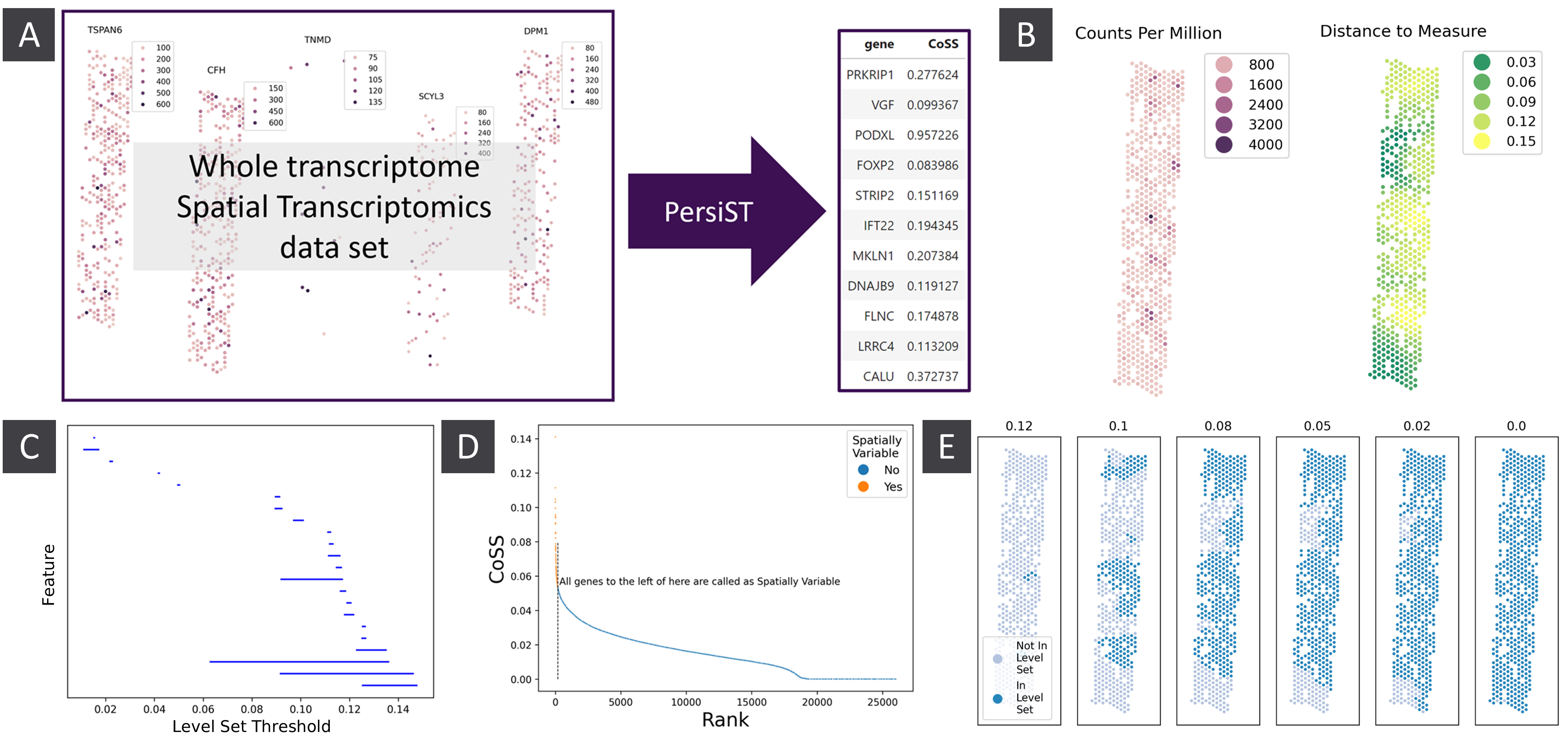}
    \caption{Computation of spatial structure scores. a) Illustration of the input and output of. We takes as input spatial transcriptomics data on some large number of genes, and output for each gene a single number quantifying the amount of spatial structure in that gene's expression. b) Original expression data (in counts per million) and the smoothed expression used for the gene PODXL in an acute kidney infection tissue sample (sample 30-10125 from \cite{kpmp_paper}). c) Barcode for the 0-dimensional persistent homology of the upper star filtration from the smoothed expression in panel b. d) CoSS-rank plot for all genes in the tissue sample in b. The CoSS cutoff for declaring a gene as spatially variable is automatically selected based on the curvature of the CoSS-Rank curve. e) Level sets of the smoothed expression in panel b, at various thresholds.}
    \label{persist_outline}
\end{figure}

Once a CoSS has been computed for each gene, SVG identification can be done by declaring all genes with a CoSS above some threshold to be spatially variable. A threshold can be automatically selected for a given sample by looking for a point of maximum curvature in the CoSS-Rank plot, where genes are ranked by their CoSS (figure \ref{persist_outline}d), but if desired any practitioner-selected cutoff can be used. Indeed, this flexibility in how permissive one wishes to be about what level of spatial structure counts as `spatially variable' is one of the advantages of the continuous spatial structure score approach to SVG identification.

\section{Methods}
In this section we fill in the technical details omitted in \emph{Method Outline}. The reader who wishes to see first the capabilities of the CoSS score when applied to real data may skip this section and proceed straight to \emph{Results}.

Mathematically, the output of a spatial transcriptomics experiment can be modelled as a collection of weighted point clouds, one for each gene. The data for a single sample consists of the co-ordinates $(x_i, y_i)_{i=1:n_\text{wells}}$ of each well, and for each gene $g$ a sequence of weights $(w_i)_{i=1:n_{\text{wells}}}$, where $w_i$ is the expression of $g$ in well $i$. Here we restrict our attention to well-based spatial transcriptomics data, in which the wells lie on a regular hexagonal or square network structure (supplementary figure 2a,b). See \emph{Supplementary Methods} for how we automatically align the given well co-ordinates to a network structure. 

To each of these weighted point clouds we wish to associate a number measuring the amount of spatial structure therein. As described in \emph{Method Outline}, we do this by computing the 0-dimensional persistent homology of the upper star filtration of each weighted point cloud, then taking the $L^2$ norm of the resulting barcode.

Whilst the underlying idea is straight-forward, there are a number of features of the problem that necessitate some more sophisticated modifications to this pipeline. 

Most significantly, the tissue slices themselves often have non-trivial spatial structure (figure~\ref{methods_figure}a) , and without correcting for this our quantification is liable to be sensitive to this. We want to ensure our methodology is robust to variations in tissue morphology, both to ensure that we are detecting genuine structure in gene expression, and to enable our metric to be used for comparison of spatial structure between different tissue samples. 

Additionally, modern well based spatial transcriptomics experiments typically output data for tens of thousands of genes. This means that our methodology will need to be computationally light, and produce a spatial structure score that can be used ``as is", without any detailed inspection of the persistent homology outputs. 

Finally, as mentioned above, we need our score to be robust to reasonable amounts of noise. 

\subsection*{Smoothed Expression}
Most of these issues can be dealt with by applying a suitable smoothing function to each weighted point cloud. Most obviously, applying a smoothing function increases robustness to small amounts of noise and trivial variations in expression from well to well. By picking a suitable smoothing function, we are also able to build in robustness to variations in tissue morphology.

The smoothing function we use is a modified form of the distance to measure of a point cloud \cite{wasserman}. For a probability density $\rho$ on $\R^2$, the distance to measure of a point $p \in \R^2$ is defined in \cite{dtm_paper} as 

\[ \dtm(p;m) = \frac{1}{m} \int_0^m \delta_a^2(x) \mathrm{d}a \]

for some pre-defined $m \in (0,1)$, where $\delta_a(x) = \inf\{r>0 : \mathbb{P}_{\rho}(B(x,r)) > a\}$ is the minimal radius of a ball around $x$ covering at least $a$ of the mass of $\rho$.

In \cite{dtm_paper} it is shown that if $\rho$ is the empirical density of a point cloud $(x_i)_{i=1:n} \subseteq \R^2$, the distance to measure is given by 

\begin{equation}\label{dtm_uniform}
    \dtm(p;m) = \frac{1}{k} \sum_{i=1}^k ||p - x^{(i)}||^2
\end{equation}

where $k = \lfloor mn \rfloor$, and $x^{(i)}$ is the $i^{th}$ nearest point to $p$ \cite{wasserman} (we will be evaluating dtm at each well, so $p = x_i$, $x^{(1)} = p$, and $||p - x^{(1)}|| = 0$). This is the average squared distance from $p$ to its $k$ nearest neighbours, where $k$ is minimal such that the combined mass on the neighbours is at least $m$. Note that the definition of $\dtm(p;m)$ depends on the co-ordinates of all points in the point cloud, but we suppress this in the notation. 

\begin{figure}
    \centering
    \includegraphics[width=0.9\textwidth]{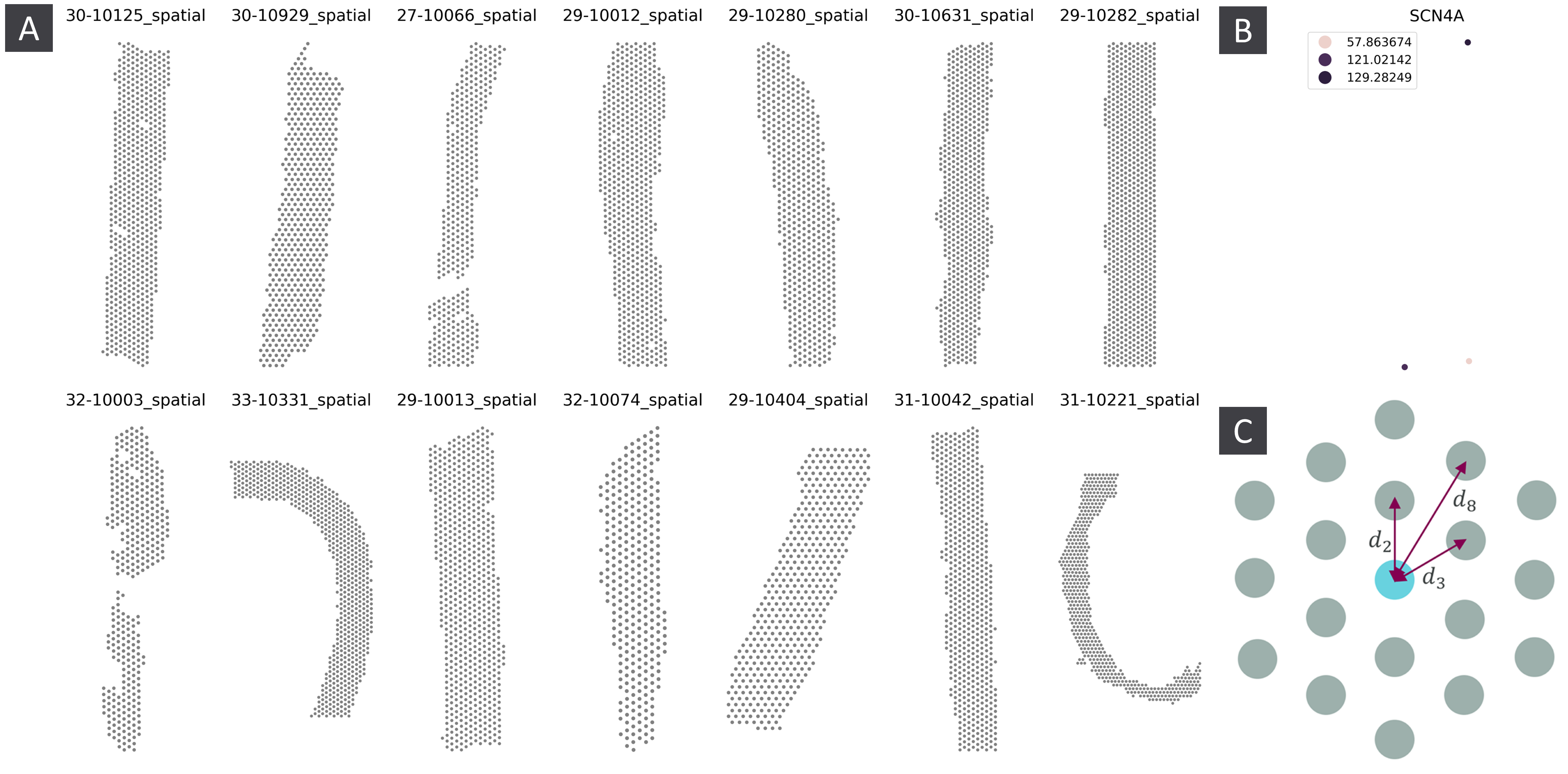}
    \caption{a) Unweighted point clouds showing the location of wells for each of the kpmp samples. b) Example of a gene with low read depth, with expression only detected in two wells. c) Distances from a node to its $2^{nd}$, $3^{rd}$ and $8^{th}$ nearest neighbours in a hexagonal grid.}
    \label{methods_figure}
\end{figure}

The distance to measure preserves many desirable features of a classic density estimator, but is more robust to noise in the input data, and has been observed to be more robust when used as an input to persistent homology \cite{wasserman}. Additionally, by using distance to measure instead of a standard kernel density estimator, we avoid over-scoring samples such as figure~\ref{methods_figure}b consisting of only a small number of wells with measured expression. These can be fairly common in spatial transcriptomics experiments with low read depth, and it is useful to be able to exclude them from any list of SVGs. 

Equation \eqref{dtm_uniform} naturally extends to the case of a weighted point cloud. Instead of summing over a constant number of nearest neighbours, we let the upper limit of the sum be $k = \min\left\{ N : \sum_{i=1}^N  w^{(i)} \geq m\right\}$, which remains the number of nearest neighbour wells needed to reach a combined mass of $m$, but will now vary depending on $p$. 

As is, the described smoothing function is highly sensitive to tissue morphology. For wells near the edge or near holes in the tissue, the sequence of distances $(||p - x^{(i)}||)_i$ will increase more quickly than for wells in the bulk of the tissue, leading to artificially higher values of $\dtm$ at these points. This is an issue that any smoothing function that smooths by looking at local expression is likely to encounter. 

We can control for this by replacing $||p - x^{(i)}||$ with $d_i$, where $d_i$ is the distance from any given vertex to its $i^{\textrm{th}}$ nearest neighbour vertex in an infinite network of the same type as the data, with $d_1 = 0$ (figure~\ref{methods_figure}c). By using network distances, we treat every well like it is in the bulk of the tissue, effectively re-arranging the wells near $p$ needed to reach a mass of $m$ such that they surround $p$ in a regular network structure, and computing $\dtm(m;p)$ as if this were what the actual data looked like.

Thus the smoothing function we end up using is given by 

\begin{equation}\label{dtm_final}
    \dtm(p;m) = \frac{1}{k} \sum_{i=1}^k d_i^2
\end{equation}

where $k = \min\left\{ N : \sum_{i=1}^N  w^{(i)} \geq m\right\}$.

Whenever we refer to distance to measure, smoothed expression, or $\dtm(p;m)$ below, we mean the quantity defined in (\ref{dtm_final}).

\subsection*{Computation of CoSS Values}
This section assumes a basic understanding of persistent homology. For a brief introduction to persistent homology, see \cite{ph_roadmap}.

We now have a collection of wells with co-ordinates $(x_i)_i$ and smoothed expression values $\dtm(x_i;m)$, producing a smoothed surface plot of the gene's expression. The smoothed expression is currently lower in regions of higher original expression, so we first invert this surface about its median, replacing $\dtm(x_i;m)$ with $z_i = \max\{\dtm(x_j;m)\}_{j= 1:n} - \dtm(x_i;m)$. This is to fit our intuition that the smoothed expression should be higher where the original expression is higher, and to integrate better with pre-existing persistent homology workflows. 

From this surface, we construct a filtered simplicial complex $(S, f)$. The nodes of $S$ are the wells $[i]$, and the edges are the $[i,j]$ where wells $i$ and $j$ are adjacent. The index for $[i]$ is $z_i$, and the index for $[i,j]$ is $\min\{z_i, z_j\}$.

Let $H$ be the 0 dimensional persistent homology of the upper star filtration on $S$. The \emph{Coefficient of Spatial Structure} (CoSS) is computed as the $L^p$ norm of the barcode of $H$. By default $p = 2$, but this can be altered by the user. A higher $p$ biases the CoSS to genes with a smaller number of regions with expression much higher than that of the surrounding tissue.

We also compute a \emph{ratio} statistic, namely the ratio of the $L^{\infty}$ to the $L^0$ norm of the barcode of $H$. This measures how much of the spatial structure in a gene's expression may be explained by a single feature. A sufficiently high value may be indicative of technical artifacts, (\emph{supplementary methods}, supplementary figure 2c,d).

\subsection*{SVG Calling}
We now have a continuous measure of the spatial structure of the observed expression of each gene. The developed measure of spatial heterogeneity is a continuous quantity, but for many tasks it is useful to have a binary yes/no call for whether a gene is spatially variable. 

We compute this by ranking all genes from highest to lowest CoSS, and looking for an ``elbow point" in the plot of CoSS versus rank for each gene (figure \ref{persist_outline}e). All genes with rank below this cutoff $K$ are declared as SV. 

We compute $K$ by looking for the point of maximal curvature in the CoSS-rank plot, using an implementation of the \emph{kneedle} algorithm \cite{kneedle_algorithm, kneed}.

\section{Results}
We evaluate the capabilities of the CoSS for detecting spatial structure in gene expression on two public Visium spatial transcriptomics data sets \cite{kpmp_paper, cardiac_paper}. These data sets were chosen as they both contain multiple samples of varying, well-defined disease phenotypes, and within each data set all samples were collected using the same data generation protocols. The two data sets also represent two ends of the spatial transcriptomics spectrum with respect to data complexity; \cite{kpmp_paper} contains samples with highly variant and quite complex morphology, where expression was measured over a small number of relatively large wells, while \cite{cardiac_paper} mostly contains samples with comparatively simple tissue morphology, with gene expression resolved over a much larger number of smaller (though still supracellular) wells .

For the results in this section we provide some biological interpretation, but our main focus is on the ability of our persistent homology pipeline to automatically detect patterns of spatial structure in large data sets. We are principally interested in a) what forms of spatial structure we are able to detect, b) how consistent the results are across samples of varying morphology and size, and c) whether topology exhibits any unique capabilities for spatial structure detection in comparison to other popularly used methodologies. In particular, throughout we compare results obtained using our topology based pipeline to those obtained using  SpatialDE, SPARK-X and Sepal. These methods were chosen to represent popularly used SVG identification methods, and the diverse range of methodologies deployed for this task; SpatialDE is based on Gaussian process regression, SPARK-X uses covariance tests, and Sepal deploys mathematical models of diffusion.

\subsection{Analysis of Kidney Disease Spatial Transcriptomics Data}\label{kpmp_analysis}
We analysed data from the Kidney Precision Medicine Project (KPMP) \cite{kpmp_paper}. This data consists of 6 Acute Kidney Injury (AKI) and 8 Chronic Kidney Disease (CKD) samples, with expression data on 26027 genes resolved to $55\mu$m wells. The number of wells varies from 317 to 787 across the samples. The tissue samples display distinct morphological variation, including some with highly irregular shapes or with multiple disconnected components (figure~\ref{methods_figure}a). Such variation presents a significant challenge for producing comparable analyses between the different samples.

The number of SVGs identified in each sample based on the automatically selected CoSS-cutoff ranged from 62 to 353. There was no correlation between the number of SVGs identified and the number of wells in the sample (table \ref{kpmp_num_svgs_stats}, supplementary figure 3a). 

Table \ref{kpmp_num_svgs_stats} provides summary statistics on the number of SVGs identified by each of the comparison methods in each of the kpmp samples, plotted in supplementary figure 3a against the number of wells in each sample. Notably, 1) SpatialDE and SPARK-X, the two methods based on null hypothesis rejection, consistently call more genes as SV than our method or Sepal, the two methods based on continuous quantification of spatial structure, 2) SpatialDE and SPARK-X exhibit much more variability in the number of SVGs called, with SPARK-X exhibiting a substantial correlation between the number of SVGs and the number of wells in a sample, and 3) Sepal consistently identifies less SVGs than we do, for some samples only calling a single digit number of genes as SV.

Spatial feature identification is effectively a form of triage, reducing a large initial number of features down to a smaller number with spatial structure for further analysis. Calling an excessively high number of features as SV increases the downstream burden on the practitioner, whilst calling too few features as SV risks missing out on important biological signal. Topology appears to hit a `sweet spot' with respect to the number of features called as SV, and exhibits greater consistency in the number of features called as SV. Moreover, the use of continuous scores enables a practitioner to triage with greater fidelity the features they wish to analyse further, by varying the score cutoff for a feature to qualify as SV.

\begin{table}[]
    \begin{center}
    \begin{tabular}{rlllllll}
    \hline
            &     & &     & & standard & & correlation with \\
     method & min & & max & & deviation & & number of wells \\
     \hline
     \vspace{1mm}
     Topology & 80 & & 332 & & 61.0 & & -0.11 \\
     \vspace{1mm}
     Sepal & 4 & & 93 & & 27.3 & & 0.21 \\
     \vspace{1mm}
     SpatialDE & 104 & & 1014 & & 282.9 & & 0.20 \\
     \vspace{1mm}
     SPARK-X & 124 & & 3886 & & 1205.8 & & 0.51 \\
     \hline
    \end{tabular}

    \vspace{3mm}

    \caption{Summary statistics for the number of SVGs called in each of the kpmp samples, for each of the comparison methods. Correlations shown are spearman correlation.}
    \end{center}
    \label{kpmp_num_svgs_stats}
\end{table}

\subsubsection*{SVG Examples}\label{kpmp_svg_examples}
To illustrate the range of spatial structures we are able to detect using persistent homology, we exhibit a couple of sets of CoSS identified SVGs, presented in groups with co-localised expression patterns, detected manually using hierarchical clustering on the observed expression values of SVGs.

\subsubsection*{Co-Localised Genes Expressed in the Glomeruli}
The type of spatial structure persistent homology is most evidently able to detect consists of multiple distinct regions of high expression surrounded by a background level of lower expression. 

\begin{figure}[!t]
    \centering
    \includegraphics[width=0.9\textwidth]{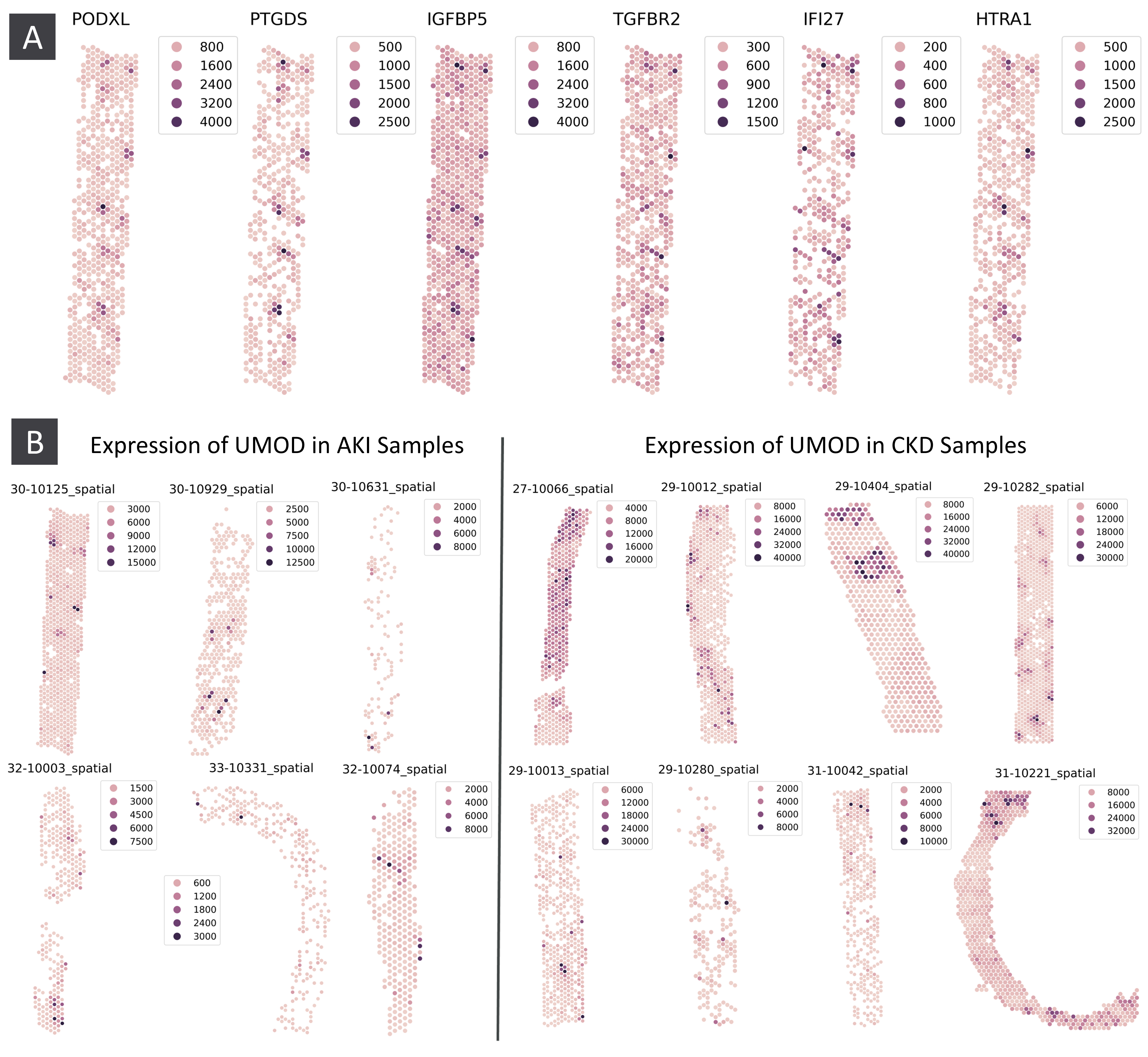}
    \caption{a) Co-expression of PODXL, PTGDS, IGFBP5, TGFBR2, IFI27, HTRA1 in kpmp sample 30-10125 (AKI) at locations corresponding to glomeruli in the tissue. Genes all identified as spatially variable by the CoSS score. b) Expression of UMOD in the AKI and CKD kpmp samples.}
    \label{kpmp_svgs}
\end{figure}

In one of the AKI samples we identified a group of SVGs highly expressed in regions of the tissue corresponding to glomeruli (figure \ref{kpmp_svgs}a), as verified by pathologist review of the accompanying H\&E images. Some of these, such as PODXL, are well-known glomerular marker genes, whilst others have not been reported as such. In particular, IFI27 is an interferon related gene, indicating the possible presence of immune activity at the glomeruli. 

SpatialDE and SPARK-X also identified the genes in figure \ref{kpmp_svgs}a as spatially variable (except SpatialDE failed to call IGFBP5 as such), but Sepal failed to identify any of these genes as spatially variable.

\subsubsection*{A Single Highly Expressed Well}
Another common type of spatial structure in the kpmp data set consists of a single highly expressed well against a background of much lower expression. Whilst such spatial variation is comparatively simple, it is important that such genes are correctly flagged as spatially variable by any automatic SVG identification process.

Supplementary figure 4 shows a collection of CoSS identified SVGs that are all highly expressed in the same well. 

The comparator methods struggled here. SpatialDE failed to call COX7B, RNF207 and NOC2L as SV, and SPARK-X, despite calling vastly more genes as SV than us (1161 compared to 188), failed to call RNF207 or NOC2L as SV. Sepal also failed to call any of these genes as SV. 

\subsubsection*{CoSS Scores Capture Structural Breakdown in CKD}\label{aki_ckd_comparison}
The continuous quantification of spatial structure provided by the CoSS can be used for additional analysis beyond identifying spatially variable genes.

For example, we can use spatial structure scores to detect differences in the spatial structure of a gene's expression between sample subgroups. In the kpmp data, we computed for each gene the difference in mean average CoSS between the AKI and CKD samples.

The gene with the highest mean CoSS difference between the AKI and CKD samples was the uromodulin encoding gene UMOD (a marker of kidney tubules). In the AKI samples the expression of UMOD is generally concentrated in a small number of very well-defined regions of high expression, whereas in the CKD samples the expression pattern of UMOD is much more diffuse, with less well-defined regions of high and low expression (figure \ref{kpmp_svgs}b). 

Progression of kidney disease is characterised by a general breakdown in the physical structure of the organ. Using persistent homology we can automatically detect and quantify this structural breakdown directly from the spatial transcriptomics data. 

Although they were not originally proposed for this application, we also inspected the mean differences in Sepal scores, and adjusted \textit{p}-values provided by SpatialDE and SPARK-X. In the case of the Sepal score, we consider this analysis a natural extension for continuous measures of spatial structure. In our analysis, UMOD only had the 6163\textsuperscript{th}, 6352\textsuperscript{th}, and 234\textsuperscript{th} greatest difference according to the Sepal score, SpatialDE q-value and SPARK-X adjusted p-value respectively. Moreover, those genes with the greatest mean difference in each case did not display any consistent notable difference in spatial structure between the AKI and CKD samples (supplementary figure 5).

\subsubsection*{Overlap Between Different SVG Identification Methods}
We investigated the overlap between the SVGs identified by our method, SpatialDE, SPARK-X and Sepal. Table \ref{comparison_table_30-10125} gives the number of genes mutually identified as SV by each pair of methods, in the kpmp sample shown in figure \ref{kpmp_svgs}a, along with a normalised version of this number divided by the total number of genes called as SV by either method. Our list of SVGs displays some overlap with the comparator methods, whilst identifying plenty of novel SVGs. There is also a substantial amount of difference between the established SVG identification methods, this lack of agreement between SVG identification methods has been observed previously \cite{svg_methods_benchmarking}.

The overlap numbers for the other samples are given in supplementary data 1, they do not differ qualitatively from the data for the sample presented here.

\begin{table}[]
    \begin{center}
    \begin{tabular}{rlllllll}
     \hline
               & Topology & & SpatialDE & &  Sepal  & & SPARK-X \\
     \hline
     \vspace{1mm}
     Topology  &    188        & &  58 (0.21)  & & 0        & &   154 (0.13)  \\
     \vspace{1mm}
     SpatialDE &    58 (0.21)  & &  144        & & 1 (0.01) & &   116 (0.10)  \\
     \vspace{1mm}
     Sepal     &    0          & &  1 (0.01)   & & 69       & &   0     \\
     \vspace{1mm}
     SPARK-X   &    154 (0.13) & &  116 (0.10) & & 0        & &   1157  \\
     \hline
    \end{tabular}

    \vspace{3mm}
    
    \caption{Number of genes mutually identified as spatially variable by different pairs of SVG identification methods in kpmp sample 30-10125. Values in brackets are normalised  by the number of genes called as spatially variable by either method. Diagonal entries are the number of SVGs called by each method.}
    \end{center}
    \label{comparison_table_30-10125}
\end{table}

\subsection{Applications to Myocardial Infarction Data}\label{cardiac_analysis}

\begin{figure}[t]
    \centering
    \includegraphics[width=0.9\textwidth]{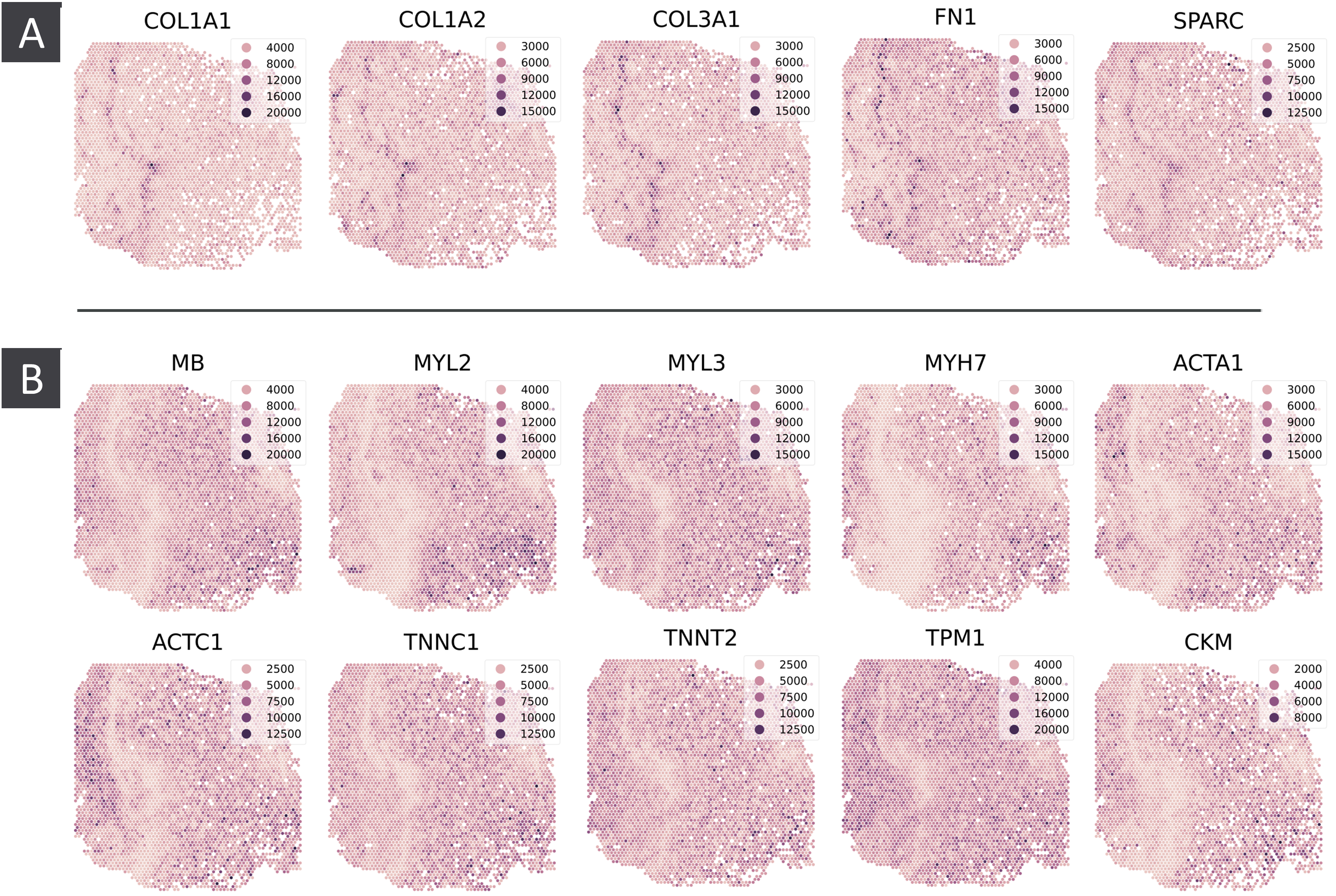}
    \caption{Select SVGs from myocardial infraction sample AKK003-157775 (Ischaemic Zone) - a) Co-expression of COL1A2, COL1A1, COL3A1, FN1 and SPARC. b) Co-expression of MB, MYL2, MYL3, MYH7, ACTA1, ACTC1, TNNC1, TNNT2, TPM1 and CKM. Genes in panel b display a noticeable drop in expression in the same region where the genes in panel a are more highly expressed.}
    \label{heart_svgs}
\end{figure}

To provide a more complete assessment of the capabilities of topology for SVG identification, we also used our methdology to analyse a spatial transcriptomics data set consisting of samples of multiple physiological zones of the heart from myocardial infarction and control patients \cite{cardiac_paper}. These samples are very different from the kidney samples analysed above in that gene expression is resolved to a much larger number of much smaller wells. These samples contain expression on 16272 genes resolved to between 1890 and 4659 $10\mu$m wells, with samples from the borderzone, fibrotic zone, ischaemic zone and remote zone, as well as control samples. 

Again we find persistent homology performs favourably compared to the benchmarked methods with respect to consistency in the number of SVGs called, and the correlation between the number of SVGs called and the number of wells in each sample (supplementary table 3, supplementary figure 3b). 

\subsubsection*{Spatial Transcriptomics Provides Additional Insight into Cardiac Fibrosis}
We identified five SVGs in an ischaemic zone sample all co-localised with COL1A2 (figure \ref{heart_svgs}a). COL1A2 has previously been identified as a driver of cardiac fibrosis \cite{li2014dynamic}, and in \cite{tomoseq} the authors use tomo-seq \cite{tomoseq_paper} to identify a group of genes whose differential regulation correlates with COL1A2 across an infarcted mouse heart. This group includes COL1A1, COL3A1, FN1 and SPARC; here we are able to verify the co-expression of these genes with COL1A2 at finer spatial resolution.

We additionally identified a group of SVGs which display a distinct drop in expression in the same region in which the above genes are highly expressed (figure \ref{heart_svgs}b). By using persistent homology, such ``voids" of expression also contribute positively to the CoSS score \cite{wasserman}. Using persistent homology and spatial transcriptomics, we are able to add additional insight into cardiac fibrosis beyond other methods and data modalities.

Sepal failed to identify any of the genes discussed in this section as SV. SPARK-X and SpatialDE identified all these genes as SV, but SpatialDE did so by calling every gene as SV. 

\subsection{Applications to Spatial Metabolomics Data}\label{msi_analysis}
Although we developed our pipeline as an application of topology to spatial transcriptomics data, the underlying methodology is agnostic as to the type of measurement recorded at each location. Viewed more broadly, we simply use persistent homology to quantify spatial structure in a weighted point cloud. Beyond spatial transcriptomics, there are many other biomedical data modalities that can be presented in this format. Using topology we avoid making any assumptions about the statistical distribution of gene expression data, so we experimented with applying the pipeline outlined in \emph{Methods} to detect spatial structure in other modalities, making no changes to the underlying methodology. 

In this section we analyse a spatial metabolomics (Mass Spectrometery Imaging, MSI) sample \cite{msi_review}. We find that persistent homology remains effective as an exploratory tool for highlighting metabolites with notable spatial structure, but that whilst our pipeline was robust to differences in the statistical properties of metabolite intensity data compared to gene expression data, the larger size (in terms of the number of points at which metabolite intensity was measured) of the sample we analysed necessitated a different level of smoothing. Additionally, converting the continuous CoSS values into a binary call of spatial variability was complicated by the qualitatively different CoSS-rank profile compared to the spatial transcriptomics samples (figure \ref{persist_outline}d, supplementary figure 1c). We discuss these points further in \emph{Supplementary Analysis}.

Persistent homology thus shows potential for analysing a broad range of spatially resolved biomedical data modalities, but more work is needed surrounding the auxiliary components of the persistent homology pipeline, and how they may be made robust to differences between data modalities.

\subsubsection*{Spatial Metabolomics Data}
In spatial transcriptomics, at each spatial location we measure the \emph{expression} of \emph{genes}. In MSI, we measure the \emph{intensity} of \emph{metabolites}. More specifically, once a grid of pixels on the tissue is decided, within each pixel the molecules of the tissue are ionised, and a mass spectrum is collected. Post data collection, computational software is used to select individual mass-to-charge (m/z) peaks, and the intensity of each m/z peak at each pixel is reported. It is at this point that the data is analogous to spatial transcriptomics data, where each feature has a measurement of abundance at each of a shared set of co-ordinates. We refer to the MSI features by their m/z ratios. The molecular identity of a specific m/z value can be determined by tandem MS (MS/MS) fragmentation, or by matching its intact mass to databases of known molecular masses within a certain mass error range (expressed in parts per million, ppm) \cite{msi_overview}.

For the sample we analyse in this section, mass spectrometry imaging was used to measure metabolite intensity on a fresh frozen rat testis at a spatial resolution of $40\mu$m.

\subsubsection*{Topology Highlights Metabolites Providing Insight into Spermatogenesis}
\begin{figure}[t]
    \centering
    \includegraphics[width=0.9\textwidth]{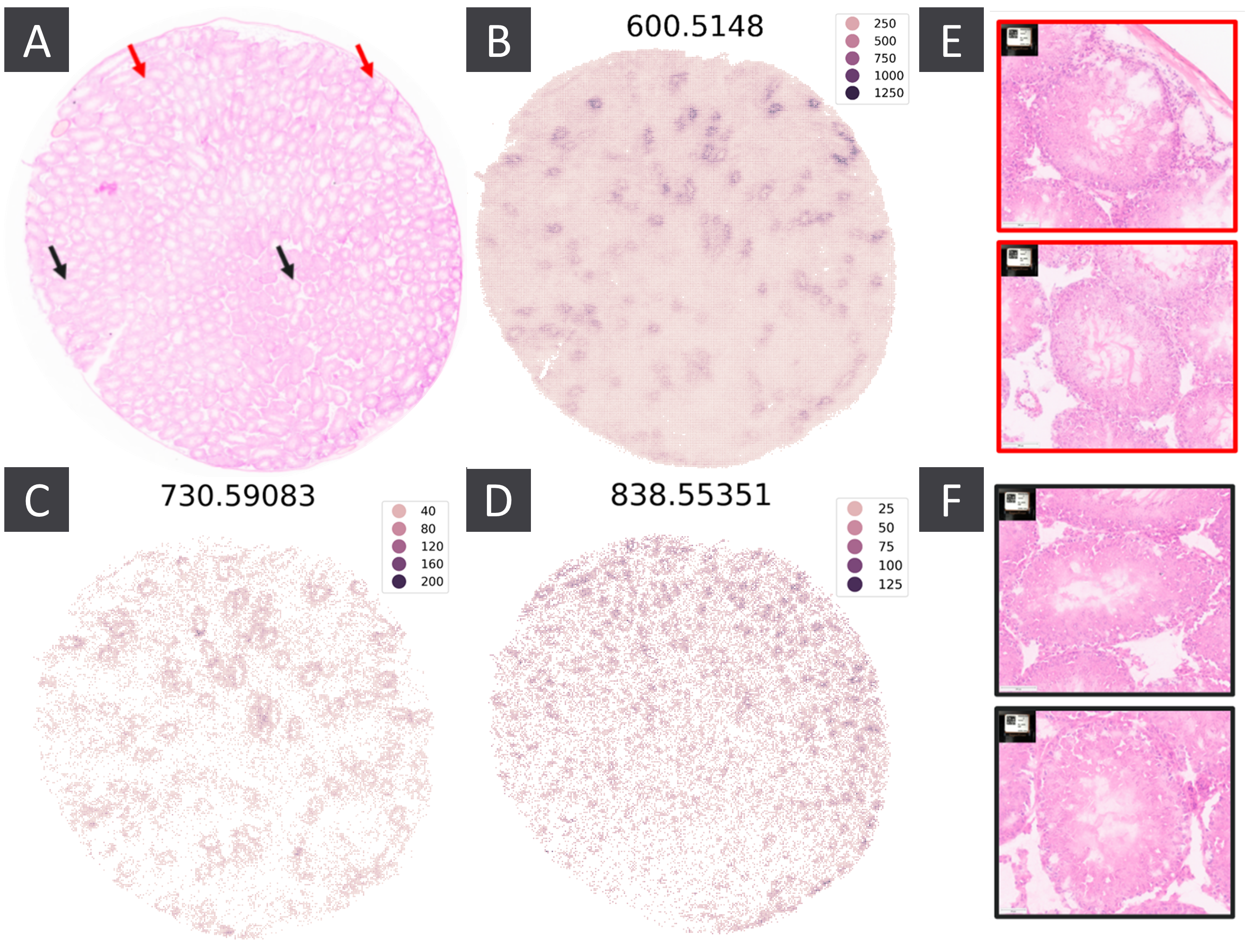}
    \caption{a) H\&E slide for the MSI sample. Red and black arrows point to example seminiferous tubules in the early/mid and late stage of spermatogenesis respectively. b,c,d) Spatial intensity of select SVMs. e,f) Zoomed in H\&E image for the regions indicated by the red and black arrows in panels a and b.}
    \label{msi_svms}
\end{figure}

Metabolites 600.5148, 602.5077, and 601.51505 were all flagged as spatially variable, and have co-localised regions of high intensity (figure \ref{msi_svms}b, supplementary figure 1a). Pathologist review of the accompanying H\&E slide (figure \ref{msi_svms}a,e,f) confirmed that these metabolites have high intensity in regions of the tissue corresponding to seminiferous tubules in the early/mid stage of spermatogenesis. Spermatogenesis occurs continuously and repeatedly in the germinal epithelium of the seminiferous tubules where these metabolites can be found. The m/z features flagged as tubular maturation markers are isotopes from the same molecular species, a ceramide identified as Cer(36:1) and detected as chloride adduct, [M+Cl]- with a mass accuracy of 3.3 ppm compared to theoretical m/z (m/z ratios identified using the Human Metabolite Database \cite{hmdb}). Levels of such sphingolipids, particularly ceramide levels, have been observed changing during the maturation phase of spermatogenesis \cite{synthase2008male}. 

Metabolite 730.59083 also shows high intensity in similar regions but is present exclusively round the edges of the tubules (figure \ref{msi_svms}c), whilst Spatially Variable Metabolites (SVMs) such as 838.55351 displayed hotspots of high intensity exclusively inside the tubules (figure \ref{msi_svms}d).  

Using persistent homology, MSI and histology data, we were able 
to identify local metabolic perturbations within our sample, and link the m/z features identified with a specific biological process. 

\section{Discussion}
We have shown how persistent homology can be used to automatically compute, from spatial omics data on a large number of features, a continuous measure of spatial structure for each feature. We have shown this measure to be a useful exploratory tool for analysing spatial transcriptomics data sets, with unique capabilities in SVG identification and differential spatial expression analysis across multiple biological settings. 

Using spatial structure scores, rather than null hypothesis rejection, better reflects the status of ``spatial variability" as a continuous descriptive property of biological systems rather than an intrinsic binary biological quantity. Moreover, we have shown that continuous spatial structure scores can be useful for multiple analytical tasks beyond just feature selection. 

By computing a spatial structure score based on principled notions of spatial structure, rather than analysing the statistical properties of gene expression data, our topology based score detected a broader range of spatial structures and performed more consistently across different biological settings. Additionally, the underlying methodology was able to produce meaningful results across multiple spatial omics modalities. 

As far as the authors are aware, our work is also a novel application of persistent homology to produce an automatic method for analysing big data, that requires no knowledge of persistent homology to be used by a practitioner. That a simple ``out of the box" application of persistent homology produced meaningful biological results for multiple spatial omics modalities indicates the broad potential of TDA for analysing complex spatial biological data.

This work highlights persistent homology as a promising tool for the automatic comparison of spatial patterns in gene expression, and in particular for the identification of spatially variable genes, with unique advantages over other commonly used methodologies in its consistency and ability to detect a broad range of spatial structures. Looking forward, a more thorough empirical and theoretical evaluation of the capabilities of persistent homology for SVG identification, a more complete analysis of the effect of different choices for auxiliary parts of the topology pipeline, such as the smoothing function, and further investigation of the use of the CoSS score for other analytical tasks would be valuable for the establishment of persistent homology as a standard out-of-the-box tool for analysing spatial omics data sets. Additionally, the past few years has seen an increasing prevalence of spatial transcriptomics experiments in which the location of each individual mRNA transcript is measured, rather than bulk expression over larger regions. An extension of the concepts discussed here to data coming in that format would enable analysis of such experiments to benefit from the capabilities of persistent homology discussed here.

Nevertheless, persistent homology remains an effective and easy to use exploratory tool for highlighting patterns in spatial transcriptomics data sets, and appears to have unique advantages compared to other popularly used methodologies with regards to the consistency of results across different biological settings, the ability to detect a much broader range of spatial structures, and to enable a wider range of analytical tasks. 

\bibliography{main} 

\end{document}